\documentclass[%
 reprint,
superscriptaddress,
 amsmath,amssymb,
 aps,
pre,
]{revtex4-2}

\usepackage{siunitx}
\usepackage{graphicx}
\usepackage{dcolumn}
\usepackage{bm}
\usepackage[export]{adjustbox}
\DeclareMathOperator\erf{\textrm{erf}}

\usepackage{xcolor}

\usepackage{tikz}
\usetikzlibrary{positioning,arrows.meta, shapes.arrows, calc}
\usetikzlibrary{external}
\begin{document}

\preprint{APS/123-QED}
\title{Creep failure in heterogeneous materials from the barrier landscape}
\author{Juan Carlos Verano-Espitia}
\email{juan-carlos.verano-espitia@univ-grenoble-alpes.fr}
\affiliation{Department of Applied Physics, Aalto University, P.O. Box 15600, 00076 Aalto, Espoo, Finland}
\affiliation{Univ. Grenoble Alpes, Univ. Savoie Mont Blanc, CNRS, IRD, Univ. Gustave Eiffel, ISTerre, 38000 Grenoble, France}
\author{Tero Mäkinen}
\affiliation{Department of Applied Physics, Aalto University, P.O. Box 15600, 00076 Aalto, Espoo, Finland}
\author{Mikko J. Alava}
\affiliation{Department of Applied Physics, Aalto University, P.O. Box 15600, 00076 Aalto, Espoo, Finland}
\affiliation{NOMATEN Centre of Excellence, National Centre for Nuclear Research, 05-400 Otwock-\'{S}wierk, Poland}
\author{Jérôme Weiss}
\affiliation{Univ. Grenoble Alpes, Univ. Savoie Mont Blanc, CNRS, IRD, Univ. Gustave Eiffel, ISTerre, 38000 Grenoble, France}

\date{\today}

\begin{abstract}

Stressed under a constant load, materials creep with a final acceleration of deformation and for any given applied stress and material, the creep failure time can strongly vary. 
We investigate creep on sheets of paper 
and confront the statistics with a simple fiber bundle model of creep failure in a disordered landscape. In the experiments, acoustic emission event times $t_j$ were recorded, and both this data and simulation event series reveal sample-dependent history effects with log-normal statistics and non-Markovian behavior. This leads to a relationship between $t_j$ and the failure time $t_f$ with a power law relationship, evolving with time. These effects and the predictability result from how the energy gap distribution develops during creep. 

\end{abstract}

\maketitle

Creep and time-dependent deformation and rupture processes under an applied constant stress are of tremendous importance in various fields, from metallurgical to civil \citep{bazant1982creep} engineering, soft matter physics \citep{cipelletti2020microscopic, leocmach2014creep, rosti2010fluctuations, koivisto2016predicting, makinen2020scale, miranda2024fractional}, to rock mechanics or geophysics \citep{duval2010creep,brantut2013rock,savage2005postseismic}. These processes result from the local deformation of soft spots due to the stress and thermal fluctuations, in line with the general picture of activated dynamics in disordered media. Central here is the concept of a local energy gap or barrier, that the system overcomes to give rise to a local excitation~\cite{cottrell1952, scholz1968creep, roux2000thermal, ciliberto2001disorder, verano2024heterogeneity, makinen2025crack, korchinski2024microscopic}. In systems with a phase transition separating an active and a frozen state, the spectrum of gaps at any given state determines the physics, in particular the dynamics at finite temperature.

Typically, creep---under constant load---starts with a decelerating primary or transient creep, followed often by a stage of constant strain rate $\dot{\varepsilon}_{min}$, and ending by an accelerating tertiary creep \citep{andrade1910metals, cottrell1952, scholz1968creep, makinen2020scale} leading to either a macroscopic rupture, or fluidization in soft matter \citep{siebenburger2012creep}. One overarching goal, especially in civil or mechanical engineering, would be to be able to predict the time of failure $t_f$ under a given load for a specific  structure, e.g. a correlation between the failure time and the minimum strain-rate has been empirically noticed for a long time \cite{monkman1957creep} and most recently a correlation between the failure time and the transtition time between primary and tertiary creep \citep{nechad2005creep, koivisto2016predicting}. Additional to monitoring the force and displacement during creep experiments, one useful way to follow the microcracking evolution in heterogeneous materials is the acoustic emission \cite{mogi1962earthquake, mogi1995earthquake, sornette2002material,rosti2010ae,viitanen2019creep}. A connection to statistical physics may be made noticing that in creep the importance of thermally activated processes has been discussed for a long time, e.g. Ref.~\citep{cottrell1952}, with suggestions of a creep lifetime in the form of an Arrhenius relation \citep{scholz1968creep, zhurkov1965solid, pomeau2002brittle, guarino2002fracture, brantut2013rock}:
\begin{equation} \label{eq:arrhenius}
t_f \sim \frac{1}{\Omega_0} \exp{\left( \frac{E}{k_BT}  \right)}
\end{equation}
where $E$ is an activation energy, $\Omega_0$ an attempt frequency \citep{zhurkov1965solid, pomeau2002brittle},  $k_B$ the Boltzmann constant and $T$ an effective temperature. The activation energy might be connected to the gap $\Delta\sigma$ between a stress threshold, or "strength", and the applied stress, via $E \sim  \Delta\sigma^{\gamma}$, where $\gamma$ is an exponent with value of $1$ in the absolute reaction rates theory proposed by Eyring \citep{eyring1936viscosity} or $3/2$ in molecular systems \cite{maloney2006barrier}. The Arrhenius form implies that small variations in strength translate into large differences in lifetime, a feature observed in heterogeneous materials ranging from paper~\cite{koivisto2016predicting} to rocks~\cite{brantut2013rock, heap2009influence, xu2012modelling, heap2011brittle} and other engineering materials~\cite{wagner1986lifetime, murthy2007stress, boumakis2019creep}.

Creep fracture in disordered media has the paradigm of so-called fiber bundle models with time-dependency that combine three key concepts: disorder in local failure resistance, re-sharing of load after an element (fiber) fails \cite{alava2006fracture, pradhan2010fbm, hansen2015fbm} and thermal activation effects \cite{roux2000thermal, ciliberto2001disorder}. Without disorder, Roux \citep{roux2000thermal} analyzed the relation between the initial and final failure times, $t_1$ and $t_f$, in the democratic (global load-sharing) fiber-bundle model (DFBM) and found $t_f \sim n^* t_1$, with $n^*$ the number of fibers needed to fail the sample with the details dependent on the applied force (impacting the gap evolution) and the temperature. 
Introducing material disorder is done through the introduction of variable fiber strengths in FBM. With large numbers of fibers, the sample-to-sample variation in e.g.~the mean strength of fibers is low, if the strengths are drawn from the same distribution. To have considerable sample-to-sample variability, the strengths need to be drawn from slightly different distributions for each sample.

In this Letter, from a sequence of acoustic emission event times $t_j$ $(j=1,2,3..)$ on creep experiments on sheets of paper, as well as from element failure times from simulations on a DFBM with thermal activation, we show that the distribution of these times over a population of samples is universally lognormal. In different creep experiments, the sequence of $t_j$ exhibits non-Markovian rank persistence. One consequence is the emergence of a power-law relationship between the event times $t_j$ and the failure time $t_f$, with an exponent that evolves with~$j$. We demonstrate that this lognormal form and rank persistence arise from a random multiplicative process, where the random multiplier also evolves with $j$. Finally, we show that the multiplicative process directly connects to the gap distribution. This allows us to reconstruct the gap properties, e.g. the minimum gap, in the experiment.

In order to avoid confusion, in what follows, $\langle X \rangle$ refers to an average over different creep experiments or FBM simulations. Whereas $\overline{X}$ refers to an arithmetic mean over the history of values, and $X_{\mathrm{ave}}$ the expected value with respect to a probability distribution---e.g.~of the exponential distribution of waiting times---in one single FBM simulation.\\

\begin{figure}[!]
    \includegraphics[width=\columnwidth]{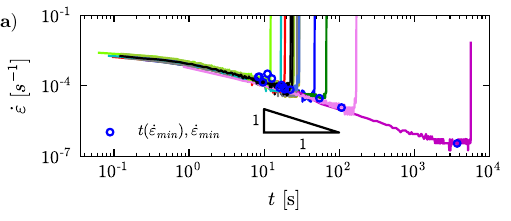}
    \includegraphics[width=\columnwidth]{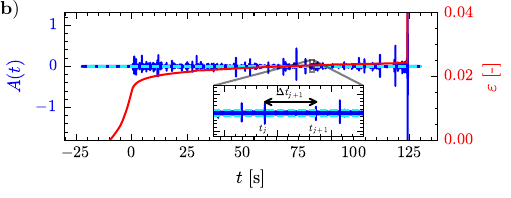}
    \caption{a) Temporal evolution of the strain rate for different experiments and identification of the minimum strain rate (blue dots). b) Amplitudes obtained from the AE sensor (blue) and evolution of the displacement (red). The cyan dashed-line is the amplitude's threshold. 
    Creep load: 0.27~\si{\kilo\newton}.} 
    \label{fig:ExpSet01}
\end{figure}

\emph{Experimental methods}---%
Tensile creep experiments in room temperature were done on 39 samples of ordinary office paper (height 100~mm, width 50~mm) where the load $F$ ($0.22~\mathrm{\si{\kilo\newton}} \leq F \leq 0.28$ \si{\kilo\newton}) is applied by an Instron Electroplus tensile machine following the procedure of Ref.~\cite{koivisto2016predicting}. To obtain relevant data, we monitored the force, the displacement (resulting in the strain rate depicted in Fig.~\ref{fig:ExpSet01}a) and also acoustic emission which was captured by a piezoelectric transducer attached to the sample to follow the microcracking \citep{rosti2010ae,viitanen2019creep}. From the continuous AE signals of amplitude $A(t)$ (arbitrary units), we define an event as the time interval the amplitude stays above an amplitude threshold, $|A(t)|>\num{1.48e-2}$, from which, we obtain event times $t_j$ $(j=1..n)$ following the procedure of Ref.~\citep{rosti2010ae}. We checked that our results are not sensitive to the choice of this threshold. Two successive events define an inter-event time $\Delta t_{j+1} = t_{j+1} - t_j$ (see Fig.~\ref{fig:ExpSet01}b). \\

\emph{Numerical methods}---%
We consider a very simple democratic fiber bundle model (DFBM) with global load sharing \citep{alava2006fracture, pradhan2010fbm} and include a small modification for increased sample-to-sample variation. Following previous works \cite{weiss2023logarithmic}, we introduce thermal activation of individual fiber ruptures through a Bortz--Kalos--Lebowitz (BKL) kinetic Monte Carlo (KMC) algorithm \cite{bortz1975KMC}. The bundle is a set of $N$ parallel fibers, with constant elastic modulus $Y$, submitted to a constant initial stress $\sigma_0=F/N$, where $F$ is the creep load. 

Fibers have initial variable strength $S_{n}$ following a Weibull distribution, commonly used for the analysis of the strength of materials \cite{weibull1939strength, alava2006fracture, wagner1984study}, and specifically in brittle materials \cite{jayatilaka1977weibull, trustrum1983weibull} such as ceramics \cite{vandenborn1991weibull}, glass fibers \cite{andersons2002weibull}, carbon fibers \cite{elasloun1989weibull, farquar1989lifetimestatistics}. To capture the variability observed between samples, we allow the scale parameter of the distribution to fluctuate slightly between simulations by setting $S_n\sim$ Weibull($\beta, \alpha + \eta_k)$, where $\beta$ is the shape parameter, $\alpha$ the scale parameter, and $\eta_k$ a zero-mean Gaussian perturbation applied to the scale parameter in the $k$th simulation. This variation reflects the heterogeneity of real materials, where e.g.~both the minimum and average fiber strengths differ from sample to sample. Accounting for such fluctuations is necessary to reproduce the wide spread of creep lifetimes observed experimentally---for example, paper sheets can differ by up to three orders of magnitude in lifetime~\cite{koivisto2016predicting}. Comparable effects have been reported in rocks~\cite{brantut2013rock}, Kevlar filaments~\citep{wagner1986lifetime}, concrete~\cite{vu2018revisiting}, and glass fibers~\cite{andersons2002weibull}, where microstructural imperfections and processing variations (e.g.~in the papermaking process~\cite{alava2006physics}) control the overall strength distribution.

In the model, the probability per unit time for a fiber $n$ to be thermally activated at the $(j+1)$th event is $p_{n,j} \sim \Omega_0 \exp{(-E_{n,j}/k_BT)}$ \cite{vineyard1957rate, kratzer2009KMC} (akin to Eq.~\ref{eq:arrhenius}), where $E_{n,j}$ is the activation energy proportional to the stress gap \cite{eyring1936viscosity, castellanos2018creep, weiss2023logarithmic, makinen2023history} and after one fiber is activated and eliminated, the time increment is exponentially distributed with a mean value of $\Delta t_{\mathrm{ave},j+1} = 1/\sum p_{n,j}$~\cite{vineyard1957rate, kratzer2009KMC, verano2024heterogeneity}. The deformation of the whole bundle takes place by the accumulation of successive thermal events that can potentially trigger avalanches of athermal ruptures of some of the surviving fibers from load redistribution. The input parameters for the FBM simulations in the following are: $N = 10000$, $T=300$, and $\sigma_0=0.2$. These parameters, especially $\sigma_0$, are chosen to prevent the individual bundles from failing immediately at the beginning of the creep stage. The fiber strength parameters are: $\alpha=1$, $\beta=5$, $\eta_k\sim\mathcal{N}(0, 0.05)$ (adding a variation of approximately $0.91 \eta_k$ to the mean and $0.21 \eta_k$ to the standard deviation of strength in individual simulations). Regarding the effect of $\eta_k$ in a population of simulations, the mean of the mean strengths over all samples is around 0.918, the standard deviation of the mean strength around 0.046 ($\sim5$~\% of the mean) and the standard deviation of the minimum strength around 0.034.
\\

\begin{figure}[!]
    \includegraphics[width=\columnwidth]{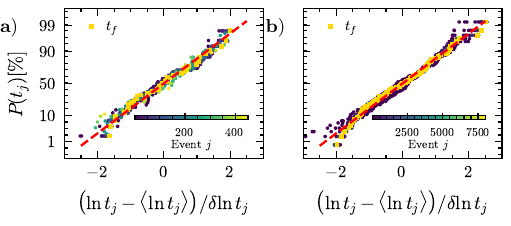}
    \includegraphics[width=\columnwidth]{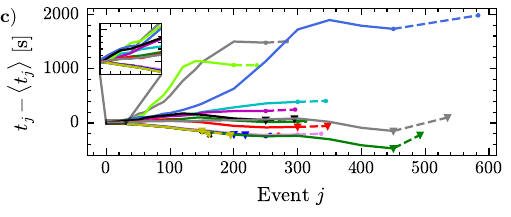}
    \includegraphics[width=\columnwidth]{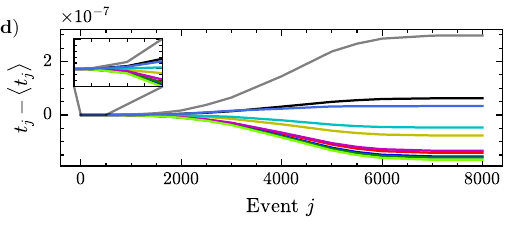}
    \caption{Universal lognormal probability distribution for the time $t_j$ in a set of a) experiments and b) FBM simulations. Both cases are represented in a normal probability plot where the x--axis represents the standard normal variable of the logarithm of $t_j$ and the y--axis are the quantiles represented as their respective cumulative probability. Non-Markovian trajectories for each individual c) experiment including $t_f$ (dashed line) and d) FBM simulation.}
    \label{fig:ExpRes01}
\end{figure}

\emph{Universal lognormal distributions and rank persistence}---%
After the AE signal is discretized into events, we
consider the probability distributions of occurrence times of AE events $t_j$ for different $j$s. We see~(Fig.~\ref{fig:ExpRes01}a) that the $P(t_j)$ follow lognormal distributions, in cumulative form
\begin{equation}
P(t_j) = \frac{1}{2} \left[ 1 + \erf{\left( \frac{\ln t_j - \langle \ln t_j \rangle}{\sqrt{2}(\delta \ln t_j)}\right)}\right],
\end{equation}
with $\langle \ln t_j \rangle$ and $\delta\ln{t_j}=\sqrt{\langle\left( \ln{t_j} - \langle \ln t_j \rangle \right)^2\rangle}$ being the mean and standard deviation of the natural logarithm of the times $t_j$. The lognormal distribution has been previously observed in the lifetime distributions of FBMs in Monte Carlo simulations~\cite{ibnabdeljalil1995timedependentfibers, mahesh2004lognormalfiber, phoenix2009timeFBM, roy2020creep}. This distribution appears already at the first events (around $j=1$). The exact same thing is true for the FBM simulations~(Fig.~\ref{fig:ExpRes01}b), except that for the very first events the distribution is not exactly lognormal. The discrepancy can easily be understood by considering that the AE monitoring may miss small events, if they are very close to the noise level of the sensor~\cite{janicevic2016interevent}. As a result, the first observed AE events might not correspond to the actual first events.

\begin{figure}[tb]
    \includegraphics[width=\columnwidth]{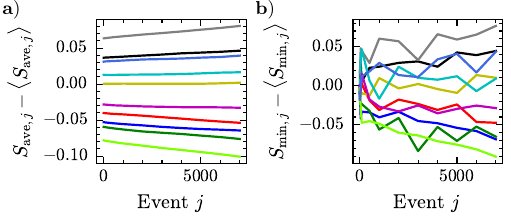}
    \caption{Non-Markovian trajectories for the a) mean and b)~minimum fiber strength in individual FBM simulations.}
    \label{fig:ExpRes01a}
\end{figure}

This universal behavior naturally leads to the question of how individual trajectories evolve over the course of an experiment: if a trajectory lies above or below the mean $\langle t_j \rangle$ at a given event index $j$, does it tend to remain there? The experimental data reveal a clear non-Markovian behavior (Fig.~\ref{fig:ExpRes01}c): trajectories that begin above or below the mean remain there for the experiment's duration. This indicates that the relative ordering of trajectories is effectively established by the earliest events. This memory effect extends even to the failure times~$t_f$ (dashed lines in Fig.~\ref{fig:ExpRes01}c). An identical pattern emerges in the FBM simulations (Fig.~\ref{fig:ExpRes01}d), confirming that this rank memory is an intrinsic feature of the underlying dynamics. The non-Markovian trajectories can also be seen in the evolution of the minimum $S_{\mathrm{min},j}$ and the average $S_{\mathrm{ave},j}$ fiber strength (see Fig.~\ref{fig:ExpRes01a}).

\begin{figure}[tb]
    \includegraphics[width=\columnwidth]{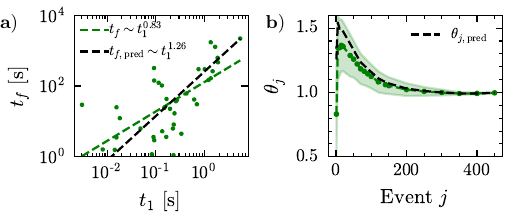}
    \includegraphics[width=\columnwidth]{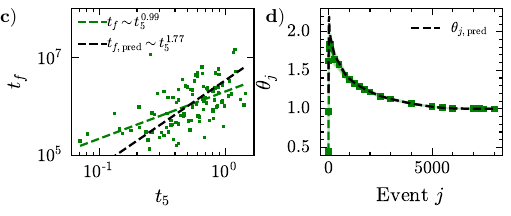}
    \caption{Creep lifetime as a function of a $jth$-event times $t_j$ in a) experiments and c) FBM simulations. The evolution of the exponent $\theta_j$ for b) the experimental data,  
     and d) the FBM results. The black dashed lines are the predicted values from Eq.~\ref{eq:t_f}.}
    \label{fig:ExpRes02}
\end{figure}

This observation of an universal lognormal distribution, and the non-Markovian rank persistence effect indicate that the relative distance of $\ln t_j$ from the mean $\langle \ln t_j \rangle$ is approximately constant (see Supplemental Material for additional analysis \cite{supplementalmaterial}). This extends up to the failure time, giving us the relation
\begin{equation}
\frac{\ln t_f - \langle \ln t_f \rangle}{\delta \ln t_f} \approx \frac{\ln t_j - \langle \ln t_j \rangle}{\delta \ln t_j},
\end{equation}
which can be reorganized to give a power-law dependency between the failure time $t_f$ and the time $t_j$
\begin{equation}
t_f \approx e^{C_{j, \mathrm{pred}}} t_j^{\theta_{j, \mathrm{pred}}},
\label{eq:t_f}
\end{equation}
where $\theta_{j, \mathrm{pred}} = \delta \ln t_f / \delta \ln t_j$, and 
$C_{j, \mathrm{pred}} = \langle \ln t_f \rangle - \theta_{j, \mathrm{pred}} \langle \ln t_j \rangle$. 
We then test this prediction by plotting $t_f$ against $t_j$ (Fig.~\ref{fig:ExpRes02}a and Supplemental Material \cite{supplementalmaterial}). The predicted exponent $\theta_{j, \mathrm{pred}}$ agrees quite well with a power-law fit to the data, particularly when accounting for the scatter in the data. Considering the exponent as a function of $j$ (Fig.~\ref{fig:ExpRes02}b), the prediction $\theta_{j, \mathrm{pred}}$ starts around $\theta_{j, \rm{pred}} = 1.3$, rises rapidly to a peak around 1.5, and then gradually drops to unity as failure is approached---consistent with the expected increase in correlation over time~\citep{monkman1957creep}. 
The fitted exponents exhibit a very similar trend: they start slightly below 1 (in line with the $t_f \sim t_1$ result from Roux~\citep{roux2000thermal}), rise to a slightly lower peak, and then drop back close to 1 near failure. 
A similar pattern emerges in the FBM simulations. The predicted exponent aligns well with the data (Fig.~\ref{fig:ExpRes02}c), and the evolution of the fitted exponent $\theta_j$ with $j$ tracks the predicted trend closely (Fig.~\ref{fig:ExpRes02}d). The primary discrepancy appears at the very first events: the prediction rises from a value around unity to a high peak but the fitted exponent values start at around 0.5 and rise to a slightly lower peak.

We verified the results presented here for different values of the Gaussian noise $\sqrt{\langle \eta_k^2 \rangle}$. We found that results match the experiments when the standard deviation of the noise is large enough, $\sqrt{\langle \eta_k^2 \rangle} > 1/\sqrt{N}$, or in other words, when the sample-to-sample variation is important. On the other hand, in the  $\sqrt{\langle \eta_k^2 \rangle} \to 0$ limit the correlation between $t_f$ and $t_j$ disappears and the behavior is clearly Markovian. See Supplemental Material for additional discussion on system size and noise effects \cite{supplementalmaterial}. \\

\begin{figure}[tb]
    \includegraphics[width=\columnwidth]{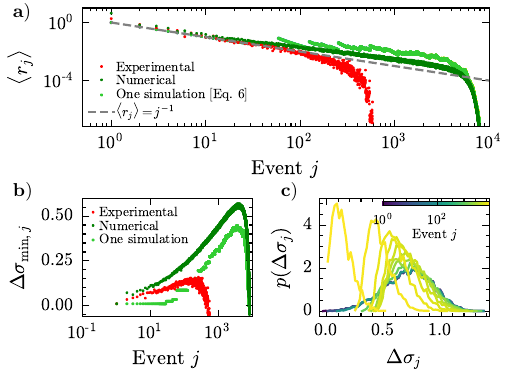}
    \caption{a) Evolution of the mean multiplier $\langle r_j \rangle = \langle \Delta t_{j+1} / t_j \rangle$ for the experimental results (red points), numerical results (green points) and theoretical result obtained from Eq.~\ref{eq002} in one FBM simulation (light green points). b) Evolution of the minimum value of the stress gap distribution for one simulation (light green points), and obtained solving $\Delta\sigma_{\mathrm{min},j}$ in Eq.~\ref{eq002} from the experimental (red points) and numerical (green points) results. c) Evolution of the stress gap distribution for different values of $j$ in one simulation. }
    \label{fig:ExpRes03}
\end{figure}

\emph{Multiplicative process and connection with the stress gap evolution}---%
To explain the observed behavior, one can consider a random multiplicative process~\cite{redner1990multiplicaive, sornette1996multiplicative}
\begin{equation}\label{eq004}
t_{j+1} = (1+r_j)t_j = \left[ \prod_{k=1}^j (1+r_k) \right]t_1, 
\end{equation}
where $r_j$ is a positive random variable, which we call the multiplier. Such process (at least with constant~$r$) produces a lognormal distribution and non-Markovian trajectories. Multiplicative processes connected to irreversible deformation have been observed in the accumulation of plastic strain in metals~\cite{tang2020lognormal, chen2021local}, wood~\cite{makinen2022wood}, and in the lognormal multiplicative cascades of sea ice deformation~\cite{marsan2004scale}. We can compute this multiplier, from $r_j = t_{j+1} / t_j - 1$, and average this over experiments or simulations (see red points for experiments and green points for simulations in Fig.~\ref{fig:ExpRes03}a).

The evolution of this random variable can be related to the evolution of the stress gap distribution (Figs.~\ref{fig:ExpRes03}b,c).
Initially the minimum gap $\Delta \sigma_{\mathrm{min}}$ is low, but as low-strength fibers are depleted, the minimum gap increases (see Fig.~\ref{fig:ExpRes03}b). This is connected to the power-law decrease in $r_j$, which can be expressed as $r_j = \Delta t_{j+1} / t_j$ where in the BKL thermal activation used in the model $\Delta t_{j+1}$ is directly connected to the gap distribution through an Arrhenius law. In an extremal dynamics approximation~\cite{sneppen1995extremal, miller1993extremal, korchinski2024microscopic}---relevant for low temperatures---the connection is dominated by the minimum gap $\Delta \sigma_{min}$ (the weakest fiber is the one eliminated after each thermal transition), i.e.~$\Delta t_{\mathrm{ave},j+1} \approx 1 / \max\{p_{n,j} \} \approx \Omega_0^{-1} \exp{(E_{\mathrm{min},j}/k_BT)}$, where $E_{\mathrm{min},j}=V\Delta\sigma_{\mathrm{min},j}=V(S_{\mathrm{min},j} - \sigma_j)$ is the minimum energy barrier to surpass in order to have a thermal transition, $V$ a constant activation volume, and $\sigma_j$ the applied global stress before the transition. In the beginning, where changes in $\Delta \sigma_{\mathrm{min}}$ are small, the waiting times are roughly constant, in which case $t_j \approx j \Delta t$, giving rise to the observed $r_j \sim j^{-1}$ behavior~(Fig.~\ref{fig:ExpRes03}a).
At a certain point the gap distribution shifts to smaller values---signifying a transition into the tertiary creep regime---and the multiplier also drops rapidly due to the decrease of $\Delta t$. Recent experiments and simulations in avalanching systems have, in primary creep, shown a power-law distribution of waiting times with a cutoff increasing linearly with time~\citep{shohat2023logarithmic, korchinski2024microscopic}, leading naturally to $\Delta t_{\mathrm{ave}} \sim t$. However, the multiplier $r$ is not directly based on the mean waiting time. We have verified that this initial $r_j \sim j^{-1}$ behavior holds even for power-law distributed $\Delta t$ with a cutoff proportional to $t$.

We can make the link to the gap distribution more concretely (see Supplemental Material for full derivation \cite{supplementalmaterial}) by noting that the event time $t_j$ is approximately equal to $j \overline{\Delta t}_j$ where $\overline{\Delta t}_j$ is the time average over the history. As stated before, assuming extremal dynamics  the mean waiting time is given by the minimum gap $\Delta t_{\mathrm{ave},j+1} \sim e^{V \Delta\sigma_{\mathrm{min}, j}/k_B T}$.
Finally, when $\overline{\Delta t_j}$ is almost constant, we can exchange arithmetic and geometric means to arrive at
\begin{equation}\label{eq002}
r_j \sim j^{-1} \exp{\left[ \frac{V \left( \Delta\sigma_{\mathrm{min},j} - \overline{\Delta\sigma_{\mathrm{min},j-1}} \right)}{k_B T} \right]}
\end{equation}
showing that the multiplier is a power-law in $j$ with a cutoff at a point given by the difference of the current minimum gap and the average of the previous minimum gaps. As can be seen from the curve based on a single numerical case in Fig.~\ref{fig:ExpRes03}a, this approximation (Eq.~\ref{eq002}) quite well recovers the observed behavior of $r_j$.

As we see the same behavior in experiments and simulations, acoustic emission monitoring then effectively becomes a direct probe of the gap distribution in the experiments. The minimum gap evolution can be extracted from the experimental data by solving $\Delta\sigma_{\mathrm{min},j}$ from Eq. \ref{eq002}, and using an iterative procedure to recover the history (see Supplemental Material for details \cite{supplementalmaterial}). The result of doing this (red points in Fig.~\ref{fig:ExpRes03}b), based on an arbitrary starting value $\Delta \sigma_{\mathrm{min}, j=0}$, shows gap evolution similar to the one seen in the simulations.\\

\emph{Summary and conclusions}---%
We have performed creep experiments on paper sheets and monitored damage accumulation through acoustic emission. The AE data were discretized into a catalog of successive event times $t_j$. We find that the distribution of $t_j$, for any $j$, over a population of samples is lognormal, and that between different samples, the sequence of $t_j$ exhibits non-Markovian rank persistence. 
This persistence enables us to derive a power-law correlation between the event times $t_j$ and the failure time $t_f$, with an exponent that evolves with $j$. The lognormal form arises from a multiplicative process, which evolves with $j$. These features can be reproduced using a minimal fiber bundle model with global load sharing, where small sample-to-sample perturbations are added to the strength distribution. The variation in the strengths is necessary to introduce sufficient sample-to-sample variation and ranking in the strength distributions, to match the experimental results. In simulations, we can directly connect the multiplicative process to the gap distribution. 

The presence of rank persistence and the resulting correlations suggest that failure times are, to some extent, predictable from early-time acoustic emission data. Remarkably, these correlations can already be detected from the very first AE events. This indicates that information about the eventual failure is embedded in the temporal structure of early damage activity.
We have focused on the correlation between individual event times $t_j$ and the final failure time $t_f$, the same methodology can be extended to examine correlations between arbitrary pairs of events $t_j$ and $t_k$ (see Supplemental Material \cite{supplementalmaterial}). This opens the door to a more general temporal inference framework, where damage history can be used to make early-stage predictions about future failure or to classify damage regimes based on their AE sequence characteristics.

The connection we establish between the multiplicative process and the gap distribution is particularly powerful. In simulations, the gap distribution is directly accessible and offers a mechanistic interpretation of the observed statistical patterns. Our results show that the evolving multiplicative process governing the $t_j$ can be understood in terms of this gap distribution, most specifically to the minimum gap. This provides a physical explanation for the lognormal form and its parameter evolution. Translating this insight back to experiments, we propose that acoustic emission data can serve as a non-invasive, indirect probe of the underlying minimum gap in real materials. That is, by analyzing AE time series, one may infer statistical properties of the damage landscape---such as the distribution and evolution of failure thresholds---even if these cannot be directly observed. This not only strengthens the bridge between statistical modeling and experimental data but also suggests other strategies for material diagnostics and failure forecasting.\\

\emph{Acknowledgments}---%
ISTerre is part of Labex OSUG@2020. This work has been supported by the French National Research Agency in the framework of the "Investissements d'Avenir" program (ANR-15-IDEX-02). 
J.C.V.E, T.M. and M.J.A. acknowledge the support from FinnCERES flagship (grant no. 151830423), Business Finland (grant nos. 211835, 211909, and 211989), and the Research Council of Finland (13361245). 
M.J.A. acknowledges support from the Academy of Finland Center of Excellence program (program nos. 278367 and 317464), as well as the Finnish Cultural Foundation. 
M.J.A. acknowledges support from the European Union Horizon 2020 research and innovation program under Grant Agreement No. 857470 and the European Regional Development Fund via the Foundation for Polish Science International Research Agenda PLUS program under Grant No. MAB PLUS/2018/8.
The authors acknowledge the computational resources provided by the Aalto University School of Science “Science-IT” project.


\bibliography{apssamp}

\end{document}


\preprint{APS/123-QED}

\title{Supplemental material: Creep failure in heterogeneous materials from the barrier landscape}


\author{Juan Carlos Verano-Espitia}
\email{juan-carlos.verano-espitia@univ-grenoble-alpes.fr}
\affiliation{Department of Applied Physics, Aalto University, P.O. Box 15600, 00076 Aalto, Espoo, Finland}
\affiliation{Univ. Grenoble Alpes, Univ. Savoie Mont Blanc, CNRS, IRD, Univ. Gustave Eiffel, ISTerre, 38000 Grenoble, France}
\author{Tero Mäkinen}
\affiliation{Department of Applied Physics, Aalto University, P.O. Box 15600, 00076 Aalto, Espoo, Finland}
\author{Mikko J. Alava}
\affiliation{Department of Applied Physics, Aalto University, P.O. Box 15600, 00076 Aalto, Espoo, Finland}
\affiliation{NOMATEN Centre of Excellence, National Centre for Nuclear Research, 05-400 Otwock-\'{S}wierk, Poland}
\author{Jérôme Weiss}
\affiliation{Univ. Grenoble Alpes, Univ. Savoie Mont Blanc, CNRS, IRD, Univ. Gustave Eiffel, ISTerre, 38000 Grenoble, France}

\date{\today}

\begin{abstract}

\end{abstract}

\maketitle


\section{Power law relationship $t_f\sim t_j^{\theta_j}$ for the experimental data}

\begin{figure}[tb]
    \includegraphics[width=\columnwidth]{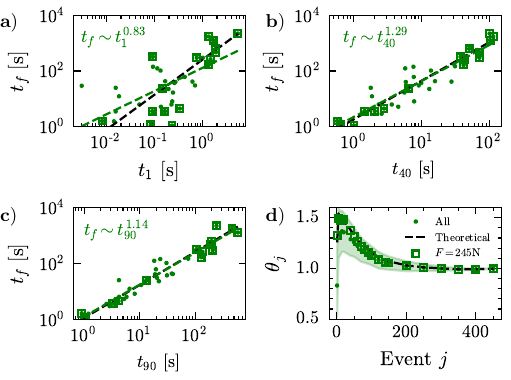}
    \caption{Creep lifetime $t_f$ as a function of different $j$th-event times in a set of 39 experiments for variable load, $0.22 \leq F \leq 0.28$ \si{\kilo\newton} (green dots) and 16 experiments submitted to a specific applied load, $F = \SI{0.245}{\kilo\newton}$ (green squares): a)~$j=1$, b)~$j=40$, c)~$j=90$. The green line is a fitted power law relation $t_f \sim t_j^{\theta_j}$ and the black line is the theoretical prediction based on the distribution of $t_j$. d) Evolution of the power law exponent $\theta_j$ with $j$ for all of the experiments and for the specific load.}
    \label{fig:Sup002}
\end{figure}

Fig. \ref{fig:Sup002} illustrates the power law evolution of the creep lifetime $t_f$ as a function of the arrival time after different $j$th event, $t_j$ (for $j=1$, 40, and 90). We can see that the experimental data follows this behavior for all of these $j$s. The behavior is also the same when one considers all the experiments (green points) or when one focuses on a very narrow load window (green squares). The evolution of the exponent $\theta$ with $j$ also does not change (green points and squares in Fig.~\ref{fig:Sup002}d). Additionally, we include the theoretical evolution of the exponent $\theta$ based on the distributions of $t_j$ (black dashed line).

\begin{figure}[tb]
    \includegraphics[width=\columnwidth]{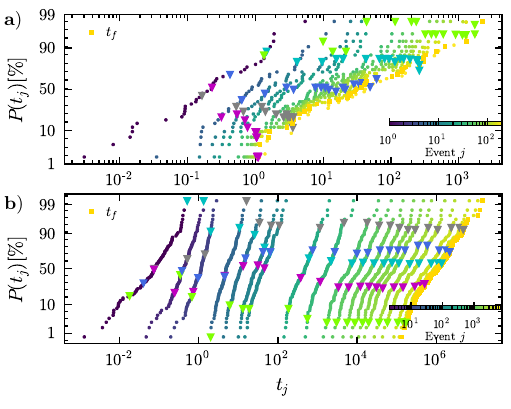}
    \caption{Universal lognormal probability distribution for the times $t_j$ (for different $j$s) for a)~experiments and b)~FBM simulations. The colored triangles are individual experiments or simulations, showing the rank-persistent non-Markovian trajectories. Both cases are represented in a normal probability plot where the x--axis represents the time $t_j$ and the y--axis are the quantiles represented as their respective cumulative probability.}
    \label{fig:Sup001}
\end{figure}

\section{Power law emergence from the universal lognormal distributions and rank persistence}
\begin{figure}[h]
    \includegraphics[width=\columnwidth]{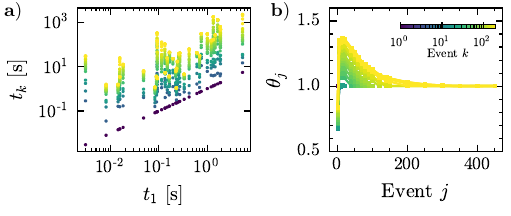}
    \includegraphics[width=\columnwidth]{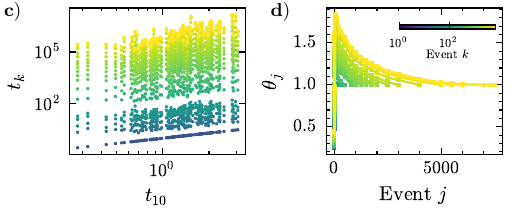}
    \caption{Times $t_{k}$ as a function of different $t_j$ in a set of a) experiments and c) FBM simulations. Evolution of the exponent $\theta_j$ as a function of $j$ for b) experiments, and d)~FBM simulations.}
    \label{fig:Sup003}
\end{figure}

In Fig.~\ref{fig:Sup001} we see the evolution of the probability distribution of times $t_j$. It follows a lognormal distribution both in the experiments (Fig.~\ref{fig:Sup001}a) and the FBM simulations (Fig.~\ref{fig:Sup001}b), from almost the first events until failure. Additionally the colored triangles represent individual experiments or simulations, indicating the rank persistence after some events have occurred ($j\approx10$ in the experimental data and $j\approx 100$ in the simulations), characteristic of non-Markovian behavior. Based on the computation done in the main paper (correlation between $t_f$ and $t_j$), it is also possible to generalize that the power law relationship, $t_f\sim t_j^{\theta_j}$, applies between any pair of times $\{t_j, t_k\}$ for all $k > j$: 
\begin{equation}
t_k \approx e^{C_{j, \mathrm{pred}}} t_j^{\theta_{j, \mathrm{pred}}},
\end{equation}
as shown in Fig.~\ref{fig:Sup003} for the experimental and numerical data.  

\begin{figure}[h]
    \centering
    \includegraphics[width=\columnwidth]{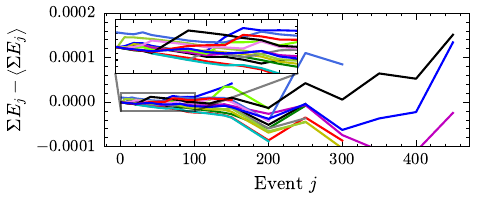}
    \caption{Markovian trajectories for the cumulative energy $\Sigma E_j$, not showing a clear ranking.}
    \label{fig:sup004}
\end{figure}

\emph{Markovian trajectories for the energies $E_j$} -- From the acoustic emission continuous signal (time and amplitude), we define an event as the time interval the amplitude stays above an amplitude threshold. For each event, we get an occurrence time $t_j$ (time of the maximum squared amplitude) and the energy $E_j$ of the event, which is the integral of the squared amplitude over the event duration \citep{rosti2010ae}. We have checked the history effects in the energies and contrary to the ocurrence times $t_j$, we do not observe a clear ranking in the sequences of cumulative energies $\Sigma E_j = \sum_{i=1}^j E_i$, i.e. no history effect (see Fig.~\ref{fig:sup004}). \\

\section{Role of the small perturbation $\eta_k$}
The initial strength of the fibers, $S_{n}$, is chosen from a Weibull distribution, commonly used for the analysis of the strength of materials \cite{weibull1939strength, alava2006fracture}, and specifically in brittle materials \cite{jayatilaka1977weibull, trustrum1983weibull, vandenborn1991weibull, andersons2002weibull, elasloun1989weibull, farquar1989lifetimestatistics}. The distribution of fiber strengths is $S_n\sim$ Weibull($\beta, \alpha + \eta_k)$, with $\alpha$ and $\beta$, the scale and the shape parameter of a Weibullian distribution. $\eta_k$ is a small zero-mean Gaussian noise implemented in the $k$th simulation, creating sample to sample variability in the minimum and average fiber strength. The other input parameters for the FBM simulations in the following are kept as in the main paper: $\alpha=1$, $\beta=5$, $T=300$, and $\sigma_0=0.2$. In order to study the effect of the noise dependency on the fiber strength distribution and the power law evolution, we used a system of $N=10000$ fibers with several values for the standard deviation $0\leq \sqrt{\langle \eta_k ^2 \rangle}  \leq 3 \times 10^{-1}$.\\

\emph{Effect on the fiber strength distribution} -- The mean of the Weibull distributed fiber strength is given by $S_{\mathrm{ave}}=(\alpha+\eta_k) \Gamma(1+1/\beta)$ (where $\Gamma$ denotes the Gamma function), so that the mean of the means over all samples is unaffected by the perturbation (as the mean of $\eta_k$ is zero), but the standard deviation over all samples has the standard deviation of the perturbation as a prefactor
\begin{equation}
\sqrt{\langle (S_{\mathrm{ave}} -\langle S_{\mathrm{ave}} \rangle )^2 \rangle} = \sqrt{\langle \eta_k ^2 \rangle} \Gamma(1+1/\beta).
\end{equation}
One can also look at the distribution of the minimum fiber strength in a bundle of $N$ fibers, which for a given value of $\eta_k$ is $S_{\rm min} \sim {\rm Weibull}(\beta, (\alpha + \eta_k) N^{-1/\beta})$. Using the law of total variance (and denoting variance with~$\mathbb{V}$), the variance in $S_{\rm min}$ between bundles is given by the conditional expected values and variances as
\begin{equation}
	\mathbb{V}\left( S_{\rm min} \right) 
    = \left\langle \mathbb{V}\left( S_{\rm min} \, | \, \eta_k \right) \right\rangle 
    + \mathbb{V}\left( \left\langle S_{\rm min} \, | \, \eta_k \right\rangle \right)
\end{equation}
so with the distributions of $S_{\rm min}$ and $\eta_k$ the standard deviation of the minimum fiber strength over all bundles becomes
\begin{equation}
\begin{aligned}
&\sqrt{\mathbb{V}\left( S_{\rm min} \right)} = 
N^{-1/\beta} \\ &\times \sqrt{\alpha^2 \left[ \Gamma\left(1+\frac{2}{\beta}\right) - \Gamma^2\left(1+\frac{1}{\beta}\right) \right] + \langle \eta_k ^2 \rangle \Gamma\left(1+\frac{2}{\beta}\right)}. 
\end{aligned}
\end{equation}
Substituting the given parameters (with $\sqrt{\langle \eta_k^2 \rangle} = 0.05$), the standard deviation of the mean values of the strength distributions of the samples is around 4.6~\%, and the minimum strength varies around 3.4~\%.

\begin{figure}[tb!]
    \centering
    \includegraphics[width=\columnwidth]{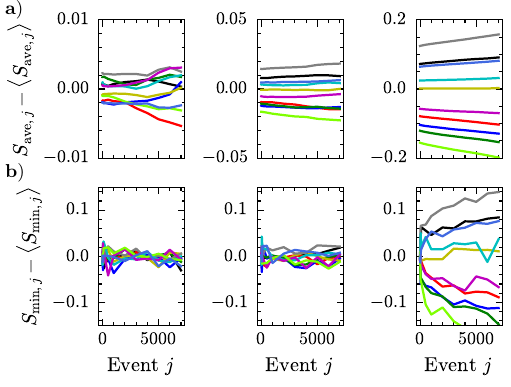}
    \caption{Effect of the standard deviation of the Gaussian noise, $\sqrt{\langle \eta_k^2 \rangle}$,  on the trajectories of a) the mean and b) the minimum value of the fiber strength distribution. The trajectories are for 10 different simulations in each case of $\sqrt{\langle \eta_k^2 \rangle}$ . From left to right: $\sqrt{\langle \eta_k^2 \rangle}=$ 0, $10^{-2}$, $10^{-1}$.}
    \label{fig:Sup005}
\end{figure}

If we choose the fiber strengths from the same distribution, i.e.~in the $\sqrt{\langle \eta_k ^2 \rangle} \to 0$ limit, the mean $S_{\mathrm{ave},j}$ and minimum $S_{\mathrm{min},j}$ remain almost invariable (see Fig.~\ref{fig:Sup005}) showing Markovian behaviour. On the other hand---as $\sqrt{\langle \eta_k ^2 \rangle}$ increases---the ranked, non-Markovian, trajectories between different samples are clearer in the mean~$S_{\mathrm{ave},j}$ (see Fig.~\ref{fig:Sup005}a) than in the minimum fiber strength~$S_{\mathrm{min},j}$ (see Fig.~\ref{fig:Sup005}b). For that reason, more variance in the distributions is needed to have observable ranking in the minimum fiber strengths (compare $\sqrt{\langle \eta_k ^2 \rangle} = 10^{-2}$ and $10^{-1}$ cases in Fig.~\ref{fig:Sup005})---which are the dominant part of the gap distribution---than just having the ranking in the means.

\begin{figure}[tb!]
    \centering
    \includegraphics[width=\columnwidth]{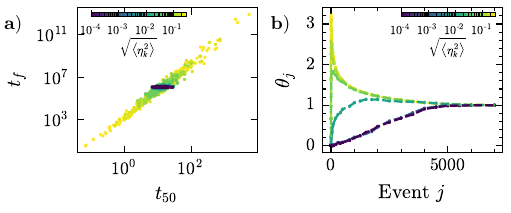}
    \includegraphics[width=\columnwidth]{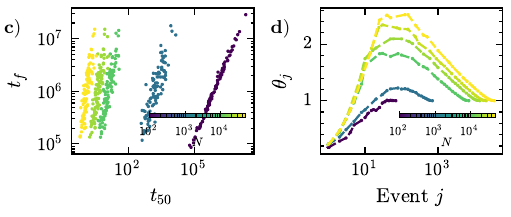}
    \caption{Creep lifetime $t_f$ as a function of time $t_j$ for a) $N=10^4$ with variable $\sqrt{\langle \eta_k^2 \rangle}$ and c) $\sqrt{\langle \eta_k^2 \rangle}=0.05$ with variable $N$. Evolution of the exponent $\theta_j$ for b) $N=10^{4}$ with variable $\sqrt{\langle \eta_k^2 \rangle}$ and d) $\sqrt{\langle \eta_k^2 \rangle}=0.05$ with variable $N$. The input parameters for the FBM simulations are: $T=300$, $\sigma_0=0.2$.}
    \label{fig:Sup011}
\end{figure}

\emph{Effect on the power law evolution} -- Fig.~\ref{fig:Sup011}a shows the failure time $t_f$ as a function of $t_j$ for simulations. We can see that as the standard deviation of the Gaussian noise increases the correlation between these two times ($t_f$ and $t_j$) becomes more evident. In the $\sqrt{\langle \eta_k^2 \rangle} \to 0$ limit, we see that for the first events the logarithmic variability around $t_j$ is much larger than the one around $t_f$, i.e. no correlation between $t_f$ and $t_j$ can be seen (see also Fig.~\ref{fig:Sup011}b for the evolution of the exponent $\theta_j$).

We also do the same analysis for the dependency of the power law evolution on the system size $N$. We evaluated several cases, $10^2 \leq N \leq 5 \times 10^4$, with constant standard deviation of the perturbation, $\sqrt{\langle \eta_k^2 \rangle}=5\times10^{-2}$. We observe that the power law dependency $t_f\sim t_j^{\theta_j}$ (shown in Fig. \ref{fig:Sup011}c for $j=50$) becomes more evident for lower values of $N$, as $t_j \to t_f$. Furthermore, the evolution of the exponent $\theta_j$ has a maximum at around the same event, $j \approx 100$, independent of $N$ (see Fig.~\ref{fig:Sup011}d). In both cases, we can see that a power law exponent $\theta_j>1$ emerges for the case $\sqrt{\langle \eta_k^2 \rangle} > 1/\sqrt{N}$.

\section{Analysis of the multiplicative random variable $r_j$}

In our thermally activated FBM the probability per unit time for an element $n$ to be thermally activated at the $(j+1)$th event is $p_{n,j} \sim \Omega_0 \exp{(-E_{n,j}/k_BT)}$ \cite{vineyard1957rate, kratzer2009KMC}, where $E_{n,j}$ is the activation energy proportional to the stress gap \cite{eyring1936viscosity, castellanos2018creep, weiss2023logarithmic, makinen2023history},
and after one fiber is activated and eliminated, the time increment is obtained from an exponential distribution with a mean of $\Delta t_{\mathrm{ave},j+1} = 1/\sum p_{n,j}$~\cite{vineyard1957rate, kratzer2009KMC, verano2024heterogeneity}. Assuming extremal dynamics \cite{sneppen1995extremal, miller1993extremal}, the time increment is dominated by the minimum stress gap (the weakest fiber is the one eliminated after each thermal transition), i.e.~$\Delta t_{\mathrm{ave},j+1} \approx 1 / \max\{p_{n,j} \} \approx \Omega_0^{-1} \exp{(E_{\mathrm{min},j}/k_BT)}$, where $E_{\mathrm{min},j}=V\Delta\sigma_{\mathrm{min},j}=V(S_{\mathrm{min},j} - \sigma_j)$ is the minimum energy barrier to surpass in order to have a thermal transition, $V$ a constant activation volume, and $\sigma_j$ the applied stress before the transition.\\

In the random multiplicative process the multiplier $r_j$ could be written as $r_j = \Delta t_{j+1} / (j\overline{\Delta t_j})$, where $\overline{\Delta t_j}$ is the mean over the previous time increments. As during the first events the variance in the waiting times is small, i.e. the waiting times are roughly constant ($\Delta t_{j+1}\approx\Delta t$), the arithmetic mean can be approximated by exchanging it with the geometric mean over the previous time increments, giving $r_j\sim j^{-1}$.
Then, assuming that the geometric mean is of the same order of magnitude of the arithmetic mean for all the remaining events (taking into account that the energy landscape has no strong variations between the values of successive weak fibers strengths), we get $\overline{\Delta t_j} \approx A_j e^{\overline{\ln{\Delta t_j}}} \approx A_j e^{\overline{\ln{ \left( \Omega_0^{-1}\exp{\left( E_{\mathrm{min}, j-1} / k_B T \right)}\right)}}}$, and the random variable $r_j$ can be rewritten as
\begin{equation}\label{eq002}
r_j \approx \frac{1}{A_j j} \exp{\left[ \frac{E_{\mathrm{min},j} - \overline{E_{\mathrm{min},j-1}}}{k_B T} \right]},
\end{equation}
where $A_j \geq 1$ is a variable that takes into account the arithmetic-geometric mean inequality and same order of magnitude between them, i.e. $1 \leq A_j < 10$, and $\overline{E_{\mathrm{min},j-1}} = V\overline{\Delta\sigma_{\mathrm{min},j-1}}=(V/j)\sum_{m=1}^{j-1} \Delta\sigma_{\mathrm{min},m}$ is the arithmetic mean of the minimum energy barrier over the previous transitions, i.e. the information of all the fibers eliminated before the new thermal transition.

Eq.~\ref{eq002} shows that the random variable $r_j$ is directly linked to the stress gap distribution, which comes from the current energy barrier to surpass and the history of the previously crossed energy barriers. Such theoretical approach is in good agreement with the results obtained for the random variable $r_j$ from the experimental data and the FBM simulations. 

\begin{figure}[tb!]
    \centering
    \includegraphics[width=\columnwidth]{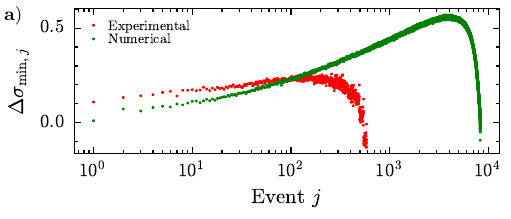}
    \includegraphics[width=\columnwidth]{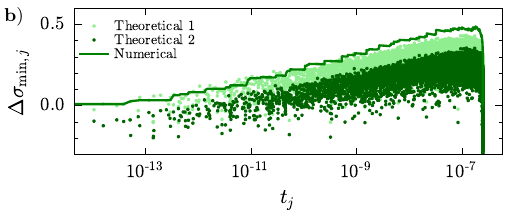}
    \caption{a) Minimum stress gap $\Delta \sigma_{\rm min, j}$ obtained from Eq.~\ref{eq005}: experimental (red points) and numerical (green points). b)~Minimum stress gap in one FBM simulation: obtained from Eq. \ref{eq005} (light green points, "Theoretical 1"), assuming $\Delta t_{j+1}=\Omega_0^{-1}\exp{\left(V\Delta\sigma_{\mathrm{min},j} / k_BT\right)}$ (dark green points, "Theoretical 2"), and the real value obtained from the output parameters of the simulation (green continuous line, "Numerical").}
    \label{fig:Sup013}
\end{figure}

In order to obtain an approximation of the variable $A_j$, one can define $\delta_j = \Delta t_j - \overline{\Delta t_j}$. If we expand $\ln{\Delta t_j}$ up to leading orders (when $\delta_j \ll \overline{\Delta t_j}$) and average, we get
$$ \overline{\ln \Delta t_j} \approx \ln \overline{\Delta t_j} + \frac{\overline{\delta_j}}{\overline{\Delta t_j}} - \frac{1}{2}\frac{\overline{\delta_j^2}}{\overline{\Delta t_j}^2} . $$
Noting that the first central moment $\overline{\delta_j}$ is zero, and denoting the square of the coefficient of variation as $c_j^2 = \overline{\delta_j^2}/ \overline{\Delta t_j}^2$, one can write the geometric mean $e^{\overline{\ln \Delta t_j}}$ in terms of the arithmetic mean $\overline{\Delta t_j}$
$$ \overline{\Delta t_j} \approx  e^{\overline{\ln \Delta t_j}} e^{ \frac{1}{2} c_j^2} $$
from which we identify $A_j=e^{c_j^2/2}$, and see that when the variance $\overline{\delta_j^2}$ is small, $c_j \to 0$, and the two means are close to equal. \\

We can use these results to obtain the minimum stress gap from the experiments. We know the value $\langle r_j \rangle$ and $\langle c_j^2 \rangle$ averaged over different experiments, as well the value of the $k_BT$ term. Solving then $\Delta \sigma_{\mathrm{min},j}$ from Eq.~\ref{eq002} yields
\begin{equation}\label{eq003}
\Delta\sigma_{\mathrm{min},j} \approx \overline{\Delta\sigma_{\mathrm{min},j-1}} + \frac{k_B T}{V} \ln{\left( j \langle r_j \rangle \right)} + \frac{1}{2}\frac{k_B T}{V} \langle c_j^2 \rangle.
\end{equation}
If we know the value of the initial gap $\Delta\sigma_{\mathrm{min},0}$, then the following gap is given by
\begin{equation*}
\Delta\sigma_{\mathrm{min},1} \approx \Delta\sigma_{\mathrm{min},0} + \frac{k_B T}{V} \ln{\left(\langle r_1 \rangle \right)} + \frac{1}{2}\frac{k_B T}{V} \langle c_1^2 \rangle.
\end{equation*}
From this point it is possible to obtain the minimum gap evolution $\Delta\sigma_{\mathrm{min},j}$ by an iterative approach, first computing $\overline{\Delta\sigma_{\mathrm{min},j-1}}$ from the history of the previous gaps and updating $\Delta\sigma_{\mathrm{min},j}$. Thus, for $j=2$ we have
\begin{equation*}
\overline{\Delta\sigma_{\mathrm{min},1}} \approx \Delta\sigma_{\mathrm{min},0} + \frac{1}{2}\frac{k_B T}{V} \ln{\left(\langle r_1 \rangle \right)} + \frac{1}{4}\frac{k_B T}{V} \langle c_1^2 \rangle
\end{equation*}
and
\begin{equation*}
\Delta\sigma_{\mathrm{min},2} \approx \overline{\Delta\sigma_{\mathrm{min},1}} + \frac{k_B T}{V} \ln{\left( 2 \langle r_2 \rangle \right)} + \frac{1}{2}\frac{k_B T}{V} \langle c_2^2 \rangle,
\end{equation*}
and we can continue this for $j=3$ and so forth.
Generalizing for any $j$, we obtain
\begin{multline}\label{eq004}
\overline{\Delta\sigma_{\mathrm{min},j-1}} \approx \Delta\sigma_{\mathrm{min},0} + \frac{k_B T}{V} \sum_{m=1}^{j-1} \frac{\ln{\left(m\langle r_m \rangle  \right)}}{m+1} \\
+ \frac{1}{2}\frac{k_B T}{V} \sum_{m=1}^{j-1} \frac{\langle c_m^2 \rangle}{m+1}\\
\approx \Delta\sigma_{\mathrm{min},0} + \frac{k_B T}{V} \ln{\left( \prod_{m=1}^{j-1} \sqrt[m+1]{ m\langle r_m \rangle} \right)}\\ + \frac{1}{2}\frac{k_B T}{V} \sum_{m=1}^{j-1} \frac{\langle c_m^2 \rangle}{m+1}.
\end{multline}
Finally, replacing this expression in Eq.~\ref{eq003}
\begin{multline}\label{eq005}
\Delta\sigma_{\mathrm{min},j} \approx \Delta\sigma_{\mathrm{min},0} \\
+ \frac{k_B T}{V} \left[ \ln{\left( \prod_{m=1}^{j-1} \sqrt[m+1]{ m\langle r_m \rangle} \right)} + \ln{\left( j \langle r_j \rangle \right)} \right]\\ 
+ \frac{1}{2}\frac{k_B T}{V} \left[ \sum_{m=1}^{j-1} \left(\frac{\langle c_m^2 \rangle}{m+1} \right)  + \langle c_j^2 \rangle \right]
\end{multline}
and choosing adequate values for the constants $V$ and $\Delta \sigma_{{\rm min},0}$, we can obtain the evolution of the minimum gap for any value of $j$ (see Fig.~\ref{fig:Sup013}a). One possible way to choose an adequate value of $\Delta \sigma_{{\rm min},0}$, could be saying that all the values of the gap should meet the criteria $\Delta \sigma_{{\rm min},j} - \overline{\Delta \sigma_{{\rm min},j-1}} \geq 0$ $\forall j$. One can see that the evolution of the term $\left( E_{\mathrm{min},j} - E_{\mathrm{min},0} \right) / k_BT$ depends only on the ocurrence times, and it is given not just by the current interevent time $\Delta t_{j+1}$ (dark green points on Fig.~\ref{fig:Sup013}b), but also by their history, ie. the average $\overline{\Delta t_j}$ and the variance $\overline{\left( \Delta t_j -  \overline{\Delta t_j} \right)^2}$ over the previous values (light green points on Fig.~\ref{fig:Sup013}b).

\begin{figure}[t!]
    \includegraphics[width=\columnwidth]{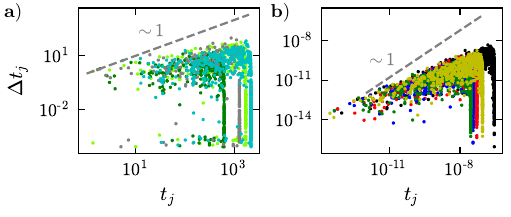}
    \caption{Waiting times $\Delta t_j$ as a function of time $t_j$ for individual a) experiments and b) numerical simulations. The dashed line shows the linear behavior $\Delta t \sim t$.}
    \label{fig:Sup012}
\end{figure}

\section{Evolution of the waiting time distribution}

Recent studies of avalanching systems in primary creep conditions---such as unfolding of crumpled Mylar  sheets~\cite{shohat2023logarithmic, korchinski2024microscopic} and corresponding simulations of bistable elastic elements~\cite{shohat2023logarithmic}
as well as elastoplastic models~\cite{korchinski2024microscopic}---have shown a power-law distribution for the waiting times
\begin{equation} \label{eq:pow_waiting_times}
    p(\Delta t) \propto t^{-1} \exp\left(-\frac{\Delta t}{\Delta t_0} \right)
\end{equation}
with a linearly time-dependent cutoff $\Delta t_0 \sim t$. This truncated power-law gives $\Delta t_{\mathrm{ave}} \sim t$, and can be seen to result in logarithmic creep behavior $\dot{\varepsilon} \sim t^{-1}$. Plotting $\Delta t_j$ vs $t_j$ in our experiments and simulations (Fig.~\ref{fig:Sup012}) shows that in our case the cutoff evolves in a power-law-like fashion, but with an exponent smaller than one.

A more important distinction of our experimental and model systems, compared to the previously mentioned cases, is the absence of Omori-like aftershock sequences in primary creep. These can be seen in the $\Delta t$ vs $t$ plot as vertical lines, spanning orders of magnitude in waiting times. In our experiments (Fig.~\ref{fig:Sup012}a) these are largely absent. Same is true for the simulations (Fig.~\ref{fig:Sup012}b), and we can see that the behavior in primary creep is dominated by single fiber failures.

With a simplistic analysis, one might arrive to a conclusion that if $\Delta t_{\mathrm{ave}}\sim t$, the multiplier $r$ would be constant. However, the multiplier comes from the statistics of subsequent events, not directly from the mean value, and even with an evolving upper envelope of the waiting time distribution, we see $r_j \sim j^{-1}$ behavior (e.g. Fig.~4 of the main paper).
We have also verified that simply sampling waiting times from a distribution given by Eq.~\ref{eq:pow_waiting_times}---where the aftershock sequences are absent---gives $r_j \sim j^{-1}$ for early times. The effect of aftershocks would only decrease the waiting times (and $r_j$), leading the behavior away from constant $r_j$.




\bibliography{apssamp}

